\documentclass[Physsubmission, Phys]{SciPost}

\hypersetup{
    colorlinks,
    linkcolor={red!50!black},
    citecolor={blue!50!black},
    urlcolor={blue!80!black}
}

\usepackage[bitstream-charter]{mathdesign}
\DeclareSymbolFont{usualmathcal}{OMS}{cmsy}{m}{n}
\DeclareSymbolFontAlphabet{\mathcal}{usualmathcal}

\usepackage{bm}

\def\kt{\bm k}
\def\ktp{\bm k^\prime}
\def\mut{\bm \mu^\prime}
\def\mutsq{\mu^{\prime2}}
\def\nn{\nonumber}

\begin{document}

\begin{center}{\Large \textbf{
Implementing Transverse Momentum Dependent splitting functions in Parton Branching evolution equations\\
}}\end{center}

\begin{center}
Lissa Keersmaekers\textsuperscript{1$\star$}
\end{center}

\begin{center}
{\bf 1} University of Antwerp
\\
* lissa.keersmaekers@uantwerpen.be
\end{center}

\begin{center}
\today
\end{center}


\definecolor{palegray}{gray}{0.95}
\begin{center}
\colorbox{palegray}{
  \begin{tabular}{rr}
  \begin{minipage}{0.1\textwidth}
    \includegraphics[width=22mm]{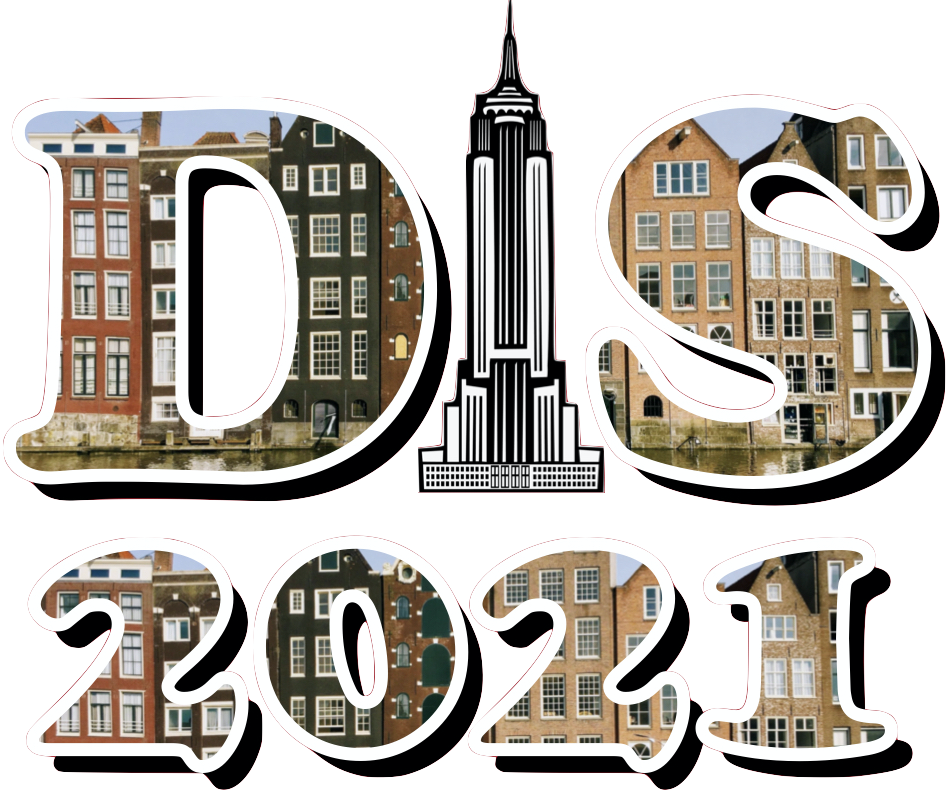}
  \end{minipage}
  &
  \begin{minipage}{0.75\textwidth}
    \begin{center}
    {\it Proceedings for the XXVIII International Workshop\\ on Deep-Inelastic Scattering and
Related Subjects,}\\
    {\it Stony Brook University, New York, USA, 12-16 April 2021} \\
    \doi{10.21468/SciPostPhysProc.?}\\
    \end{center}
  \end{minipage}
\end{tabular}
}
\end{center}

\section*{Abstract}
{\bf
The Parton Branching (PB) approach describes the evolution of transverse momentum dependent (TMD) parton densities. We propose to extend the PB method by including TMD splitting functions, instead of the DGLAP splitting functions which assume strong ordering in transverse momentum. We present the evolution equations and their numerical solution, which is the first Monte Carlo implementation including TMD splitting functions.
}


\section{Introduction}
\label{sec:intro}

TMD factorization theorems (see \cite{Angeles-Martinez:2015sea} and references therein) 
are important for precise theoretical predictions of physical observables 
 such as the Drell-Yan (DY) transverse momentum spectrum in hadronic collisions. The Parton Branching (PB) method~\cite{HAUTMANN2017446,Hautmann2018,PhysRevD.99.074008} allows one to obtain the evolution of TMD Parton Distribution Functions (TMD PDFs) in terms of Sudakov form factors, real-emission splitting functions and 
 angular-ordering phase space constraints. The method has recently been applied to  
 DY~\cite{PhysRevD.100.074027,HAUTMANN2019114795,Martinez2020} and photon-induced~\cite{JUNG2021136299} lepton-pair production, and to 
 DY + jets production~\cite{Martinez:2021chk}. It has been implemented in  the   Monte Carlo (MC) event generator  {\sc Cascade}~\cite{Baranov2021,Jung2010}.
 
Currently, the PB method uses Dokshitzer-Gribov-Lipatov-Altarelli-Parisi (DGLAP)~\cite{Gribov:1972ri,Altarelli:1977zs,Dokshitzer:1977sg}  
splitting functions, which assume that the partons along the branching decay chain are 
strongly ordered in transverse momenta. 
From high-energy factorization~\cite{Catani:1990xk,Catani:1990eg,Catani:1993ww}, it is known that 
potentially large corrections to strongly-ordered branchings arise for small longitudinal-momentum fractions $x$~\cite{Hautmann:2012pf,Hautmann:2012qr}. 
These corrections can be taken into account 
by generalizing  the concept of DGLAP splitting functions to that of TMD splitting functions~\cite{CATANI1994475,Catani:1993rn}.  
A calculational programme  of TMD splitting functions is pursued 
in~\cite{Hautmann:2012sh,Gituliar2016,Hentschinski:2016wya,Hentschinski2018,Hentschinski:2021lsh}.

We propose in this work an implementation of TMD splitting functions within the PB method. It is the first MC implementation that uses TMD splitting functions and a first step towards a new MC that includes small-$x$ physics.

\section{Evolution equations}

In this work, three different scenarios for PB evolution equations are studied~\cite{lissaetal-inprep}. The equations for momentum-weighted TMD PDFs $\tilde{\mathcal{A}}_a(x,\kt,\mu^2)=x\tilde{\mathcal{A}}_a(x,\kt,\mu^2)$ for a parton of flavour $a$, evaluated at a scale $\mu$, with $x$ the longitudinal momentum fraction of the proton and $\kt$ the transverse momentum, are given by 
\begin{align}
\tilde{\mathcal{A}}_a(x,\kt,\mu^2)=&{\Delta_a(\mu^2)}\tilde{\mathcal{A}}_a(x,\kt,\mu_0^2)\nn+\sum_b\int\frac{d^2\mut}{\pi\mutsq}\frac{\Delta_a(\mu^2)}{\Delta_a(\mutsq)}\Theta(\mu^2-\mutsq)\Theta(\mutsq-\mu_0^2)\times\\
&\times\int_x^{z_M}dz{P_{ab}}\tilde{\mathcal{A}}_b(\frac{x}{z},\kt+(1-z)\mut,\mutsq), 
\label{eq:evolution}
\end{align}
where $\Delta_a$ is the Sudakov form factor for parton $a$, $P_{ab}$ is the real-emission  function for parton splitting $b \to a$, and the phase-space angular ordering 
is embodied in i) the running coupling, ii) the  relationship between the evolution variable and the transverse momentum,  and iii) the   
 soft gluon resolution scale $z_M$. The latter  separates resolvable from non-resolvable branchings, and is taken to be dynamical, i.e.,  
 dependent on the evolution scale $\mu'$: $z_M =1-q_0/\mu'$, where 
 the parameter $q_0$ represents the minimal transverse momentum of the emitted parton. The main features of this approach 
are described in\cite{HAUTMANN2019114795}. 
The three scenarios are characterized as follows: 
\begin{enumerate}
\item $P_{ab}=P^{col}_{ab}(z,\mutsq)$ and $\Delta_a(\mu^2)=\Delta^{col}_a(\mu^2)$; 
\item $P_{ab}=P_{ab}^{TMD}(z,\ktp,\mut)$, with $\ktp=\kt+(1-z)\mut$, and $\Delta_a(\mu^2)=\Delta^{col}_a(\mu^2)$; 
\item $P_{ab}=P_{ab}^{TMD}(z,\ktp,\mut)$ and $\Delta_a(\mu^2)=\Delta_a^{TMD}(\mu^2,\kt^2)$. 
\end{enumerate}

The resolvable branching probabilities of the first condition $P^{col}_{ab}(z,\mutsq)$ are the real emission parts of the DGLAP splitting functions. The TMD splitting functions in their original variables $\tilde P_{ab}^{R}(z,\ktp,\tilde \kt)$, which can be found in \cite{Hentschinski2018}, are defined within the 2-gluon irreducible kernels, which integrate over the boost-invariant transverse momentum $\tilde \kt=\kt-z\ktp$. For the second and third condition we use the TMD splitting functions $P_{ab}^{TMD}(z,\ktp,\mut)={\frac{d^2\tilde \kt}{\tilde \kt^2}\frac{\mutsq}{d^2\mut}}\tilde P_{ab}^{R}(z,\ktp,\tilde \kt)$, were we have chosen to absorb a Jacobian in their redefinition. The collinear Sudakov form factor, which resums non-resolvable branchings and virtual corrections, is given by \\$\Delta^{col}_a(\mu^2)=\exp[-\sum_b\int_{\mu_0^2}^{\mu^2}\frac{d\mu^{\prime 2}}{\mu^{\prime 2}}\int_0^{z_M}dz\ z\  P^{col}_{ba}(z,\mu^{\prime 2})]$. In the standard PB method, which corresponds to the first condition, it has the interpretation of the probability of an evolution without any resolvable branching. Since we change the resolvable branching probabilities in the second condition, this interpretation is no longer valid. In the third condition the TMD Sudakov form factor  $\Delta_a^{TMD}(\mu^2,\kt^2)=\exp[-\sum_b\int_{\mu_0^2}^{\mu^2}\frac{d\mutsq}{\mutsq}\int_0^{2\pi}\frac{d\phi}{2\pi}\int_0^{z_M}dz\ z\  P^{TMD}_{ba}\left(z,\kt,\mut\right)],$ where $\phi$ is the angle between $\kt$ and $\mut$, is defined such that this interpretation is again valid. This condition has effects of TMD splitting functions in both resolvable and non-resolvable branchings. At $\kt=0$, the TMD Sudakov form factor is equal to the collinear Sudakov form factor. When $\kt$ increases, the TMD Sudakov form factors decrease for both gluons and quarks.\\
A property of the equations with the first and third condition is that the momentum of the proton is conserved. The second condition doesn't conserve the proton's momentum. This can be shown analytically, and is shown numerically in table \ref{tab:momentumsum}.\\
The evolution equations are here shown for TMD PDFs, but to obtain collinear PDFs, one can simply integrate over the transverse momentum $\kt$: $\tilde f_a(x,\mu^2)=\int \frac{d^2\kt}{\pi} \tilde{\mathcal{A}}_a(x,\kt,\mu^2)$.\\ The equations can be solved with MC techniques.
\section{Numerical results}
\begin{figure}
\begin{minipage}{0.5\linewidth}
\centerline{\includegraphics[width=0.7\linewidth]{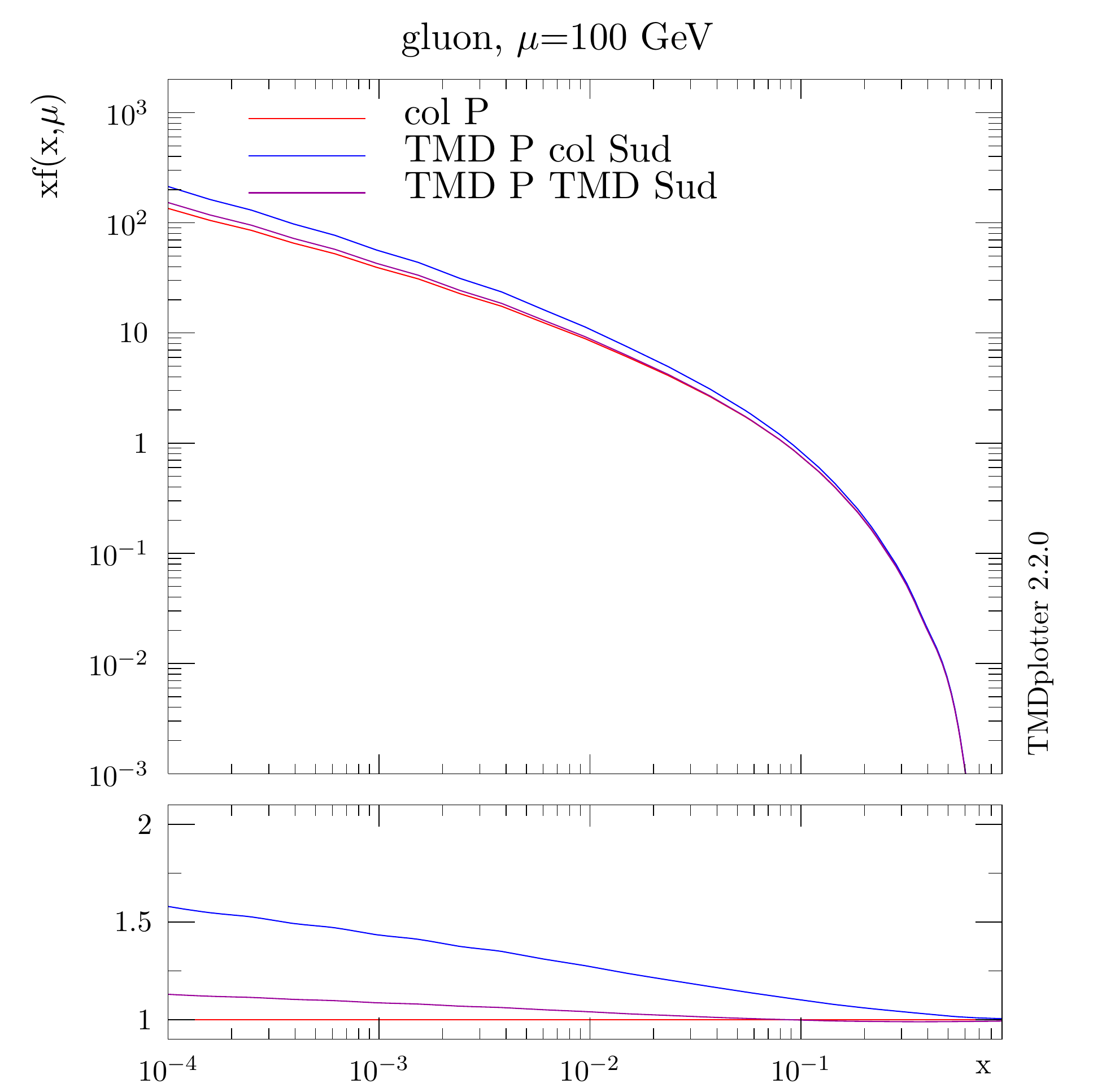}}
\end{minipage}
\hfill
\begin{minipage}{0.5\linewidth}
\centerline{\includegraphics[width=0.7\linewidth]{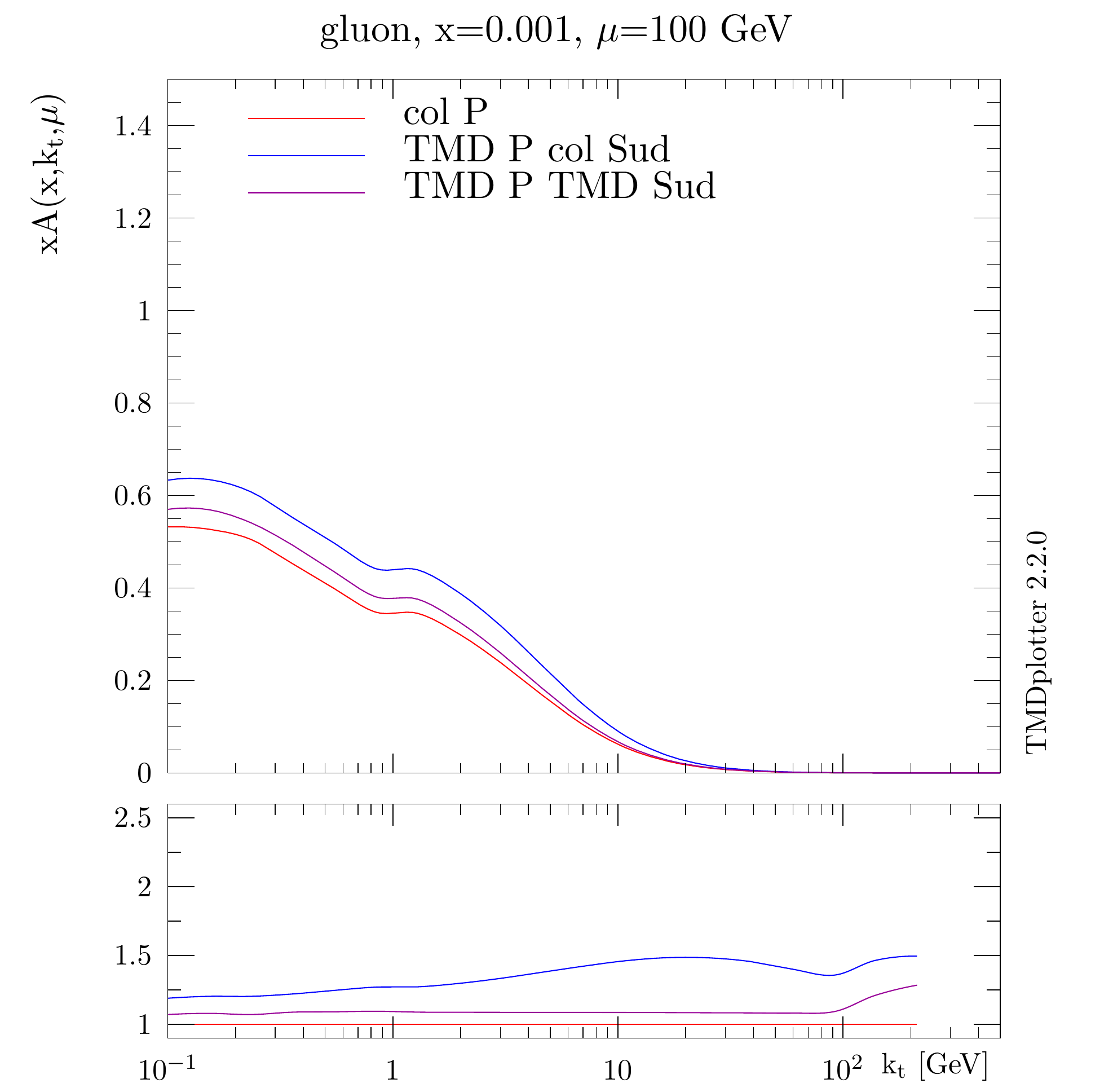}}
\end{minipage}
\caption[]{Collinear (left) and TMD (right) PDFs for gluons. The bottom plots show the ratio between the curves.}
\label{fig:PDFs}
\end{figure}

\begin{figure}
\begin{minipage}{0.5\linewidth}
\centerline{\includegraphics[width=0.7\linewidth]{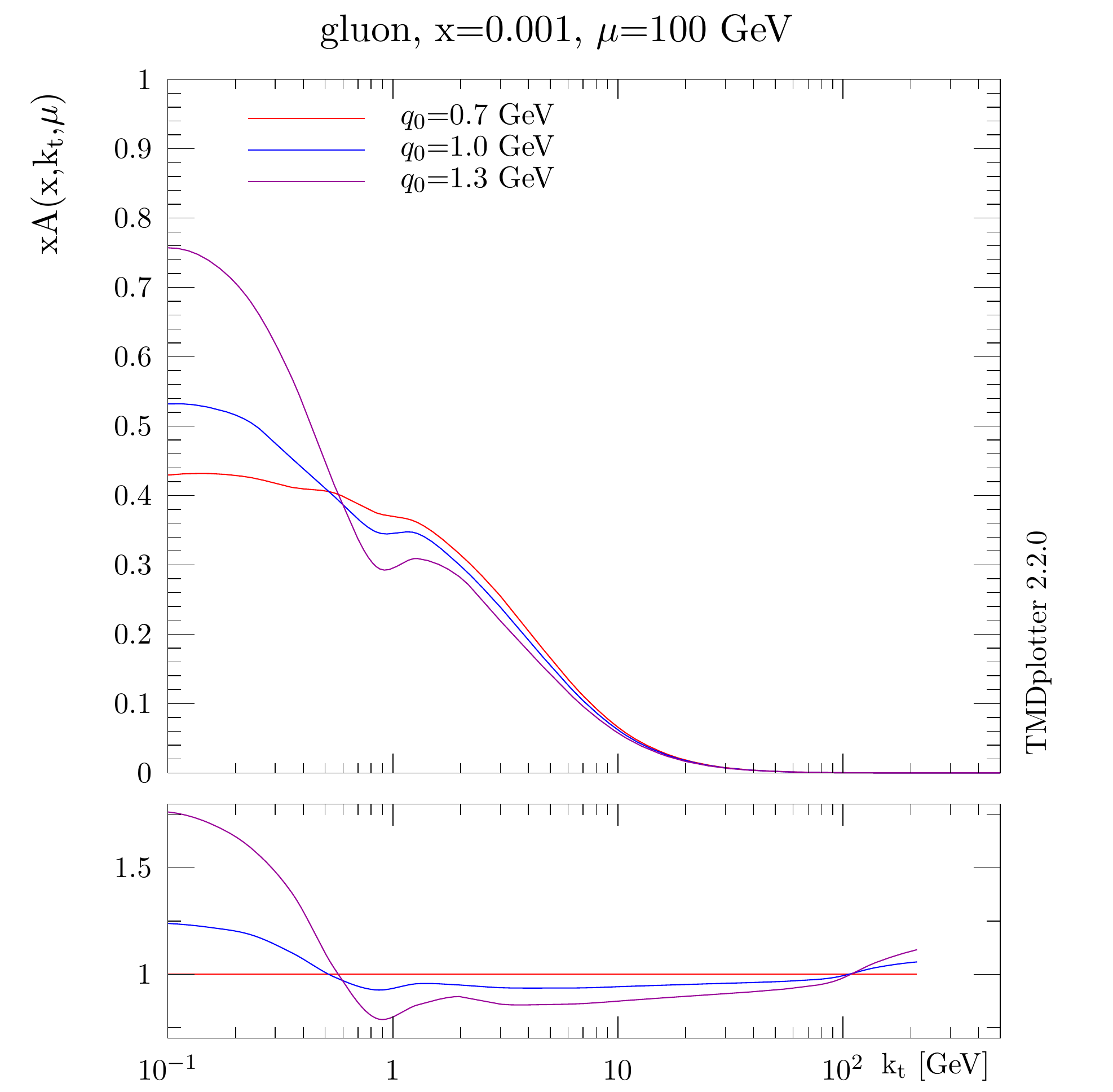}}
\end{minipage}
\hfill
\begin{minipage}{0.5\linewidth}
\centerline{\includegraphics[width=0.7\linewidth]{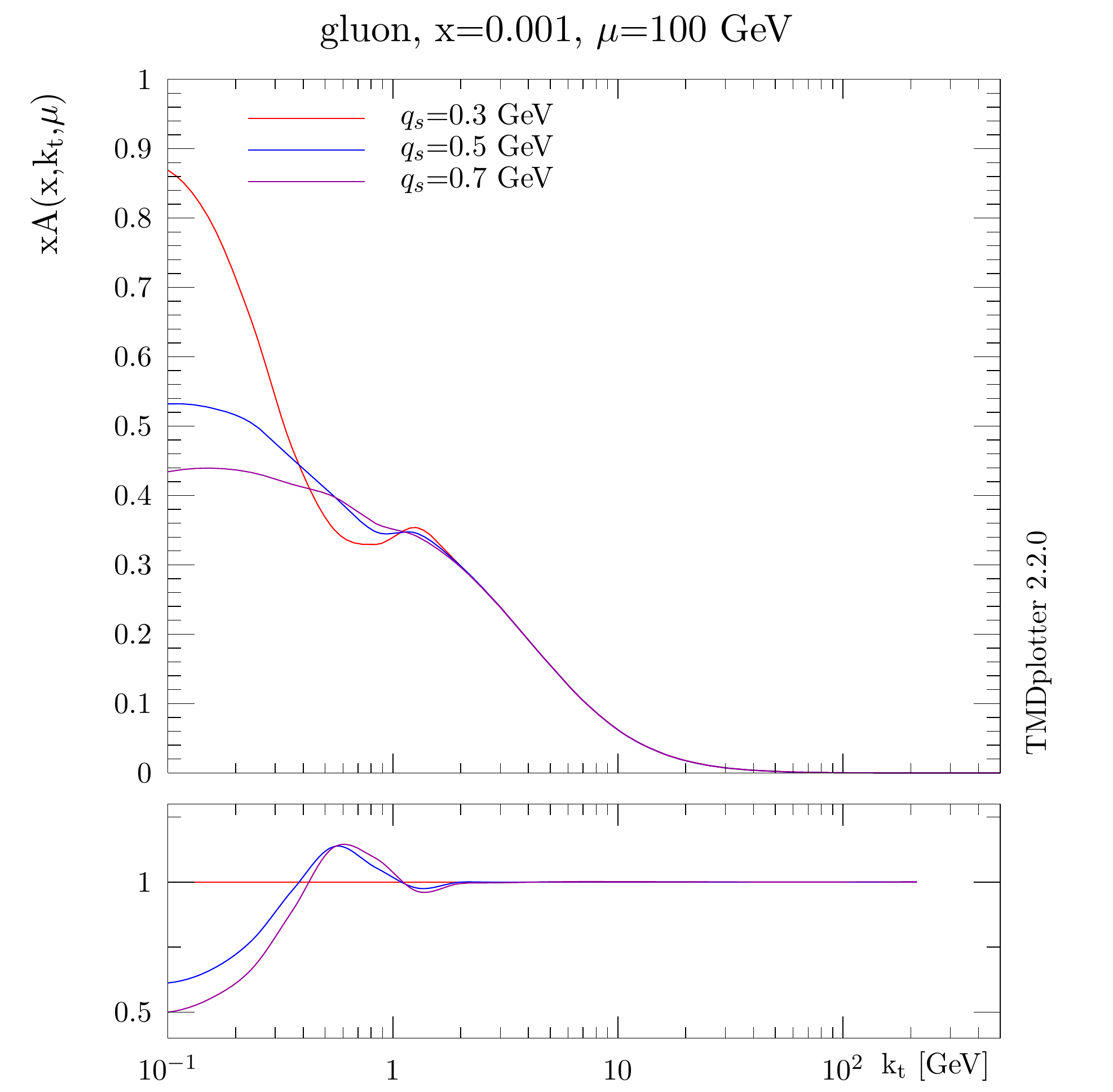}}
\end{minipage}
\caption[]{Gluon TMD PDFs for different values of the non-perturbative input $q_0$ and $q_s$. The bottom plots show the ratio between the curves.}
\label{fig:nonperturbativeinput}
\end{figure}

In this section, we show results for the evolution equations \ref{eq:evolution} obtained with the same initial parametrization for all scenarios $\tilde{\mathcal{A}}_a(x, k_{\bot,0},\mu_0^2)= x f_a(x,\mu_0^2)\cdot \frac{1}{q_s^2}\exp\left(-{ k_{\bot,0}^2/ q_s^2}\right)$ at scale $\mu_0^2=1.9\ \textrm{GeV}^2$. The x-dependence is generated according to a collinear PDF $x f_a(x,\mu_0^2)$ which is chosen to HERAPDF20\_LO\_EIG and the $k_\bot$-dependence is generated according to a Gaussian, with $q_s=0.5$ GeV. By using fixed starting distributions, we can study the impact of each element in the evolution equations. However, before the TMD PDFs can be applied to phenomenology, they should be fitted, which is left for future work.
\begin{table}[t]
\caption[]{Check of momentum conservation: $\int_{10^{-5}}^1dx\tilde f_a(x,\mu^2)$.}
\label{tab:momentumsum}
\vspace{0.4cm}
\begin{center}
\begin{tabular}{|l|c|c|c|}
\hline
$\mu^2$ (GeV)&
$P^{col}_{ab}$, $\Delta^{col}_{a}$ &
$P^{TMD}_{ab}$, $\Delta^{col}_{a}$ & $P^{TMD}_{ab}$, $\Delta^{TMD}_{a}$\\
&(condition 1) &(condition 2) &(condition 3)
\\ \hline 
10      & 0.999    & 1.007  & 0.999 \\
100     & 0.997    & 1.045  & 0.997 \\
1000    & 0.995    & 1.091  & 0.994 \\
10000   & 0.992    & 1.129  & 0.991 \\
100000  & 0.984    & 1.148  & 0.983  \\ \hline
\end{tabular}
\end{center}
\end{table}
In figure \ref{fig:PDFs}, we show the gluon distribution at scale $\mu=100$ GeV obtained with condition 1 (col P), condition 2 (TMD P col Sud) and condition 3 (TMD P TMD Sud). The bottom plots show the ratio between the different curves, compared to condition 1. With these ratio plots one can easily see the differences between the evolved TMDs in the whole $x$ or $k_\bot$ region. Left, the integrated TMD PDF versus $x$ is shown. The TMD splitting functions affect the distributions already at the level of integrated TMDs. The TMD splitting functions reduce to the DGLAP splitting functions for $\kt'\rightarrow0$ and therefore the effects of them are small in the large-x region. A suppression in the PDF can be seen for the model with TMD Sudakov form factor compared to the PDF from the model that uses TMD splitting functions and a collinear Sudakov form factor. Right, the TMD PDF versus $k_\bot$ at $x=0.001$ is shown. The whole $k_\bot$-region is affected by the TMD splitting functions. The suppression from the TMD Sudakov form factor is visible in the whole $k_\bot$-region. There are kinks visible in the $k_\bot$-spectrum due to the non-perturbative input.
\\
In figure \ref{fig:nonperturbativeinput} we show the influence of non-perturbative input. We do this for the first condition, but condition 3 gives similar results. The transverse momentum of a parton after $n$ branchings is given by $\kt=\bm k_{0}-\sum_{i=1}^n \bm q_i$, with $\bm q_i=(1-z)\bm \mu_i$ the transverse momentum of the emitted parton. Since we use the dynamical resolution scale, we have $q_{\bot,i}>q_0$ and the transverse momentum from the evolution begins to build up around the value of $q_0$. But since we have a vector sum, $k_\bot$ below $q_0$ can be reached and the peak can be smeared by many branchings. In the left figure we show the gluon TMD PDF vs $k_\bot$ for different values of $q_0$. We see that when we lower $q_0$, the curve gets smoother, this has two reasons: with low $q_0$ there are more resolvable branchings and therefore the peak from evolution is more smeared out and with low $q_0$ the overlap is larger with the peak from intrinsic $k_\bot$ around 0. In the right figure we show the effects of the width of the Gaussian function for intrinsic $k_\bot$, by varying $q_s$, while $q_0$ is fixed at 1 Gev. When $q_s$ is close to $q_0$, the curve is smoother than when $q_s$ is smaller, again because of the overlap between the two peaks. We can see that $q_s$ only affects the small $k_\bot$-region.

\section{Conclusion}

We have presented a Monte Carlo implementation of TMD splitting functions in the Parton Branching approach. The TMD splitting functions affect both resolvable branchings and Sudakov form factors. The evolution with TMD splitting functions has impact on both collinear and TMD parton distributions. The effects are visible in the small-$x$ region of 
PDFs and throughout the whole $k_\bot$-spectrum of TMD PDFs. Phenomenological studies are warranted at the LHC as well as at future hadron-hadron~\cite{Mangano:2016jyj,Azzi:2019yne} and lepton-hadron~\cite{LHeC:2020van,Proceedings:2020eah} colliders.

\section*{Acknowledgements}
I thank Francesco Hautmann, Martin Hentschinski, Hannes Jung, Aleksander Kusina, Krzysztof Kutak and Aleksandra Lelek for collaboration and discussion.
\bibliography{DIS}

\nolinenumbers

\end{document}